\documentclass[12pt]{iopart}
\usepackage{amssymb}
\usepackage{graphicx}
\usepackage{dcolumn}
\usepackage{bm}
\usepackage{epsfig}
\newcommand{\beq}{\begin{equation}}
\newcommand{\eeq}{\end{equation}}
\newcommand{\beqa}{\begin{eqnarray}}
\newcommand{\eeqa}{\end{eqnarray}}
\newcommand{\ba}{\begin{array}}
\newcommand{\ea}{\end{array}}
\usepackage{color}
\begin{document}

\title[Atomic Josephson junction with two bosonic species]
{Atomic Josephson junction with two bosonic species}
\author{Giovanni Mazzarella$^{1}$, Marco Moratti$^{1}$, Luca Salasnich$^{2}$,
Mario Salerno$^{3}$, and Flavio Toigo$^{1}$} \address{$^{1}$Dipartimento di Fisica
``Galileo Galilei'' and CNISM, Universit\`a di Padova, Via Marzolo 8, 35131
Padova, Italy
\\$^{2}$CNR-INFM and CNISM, Unit\`a di Padova, Via Marzolo 8, 35131
Padova, Italy
\\$^{3}$Dipartimento di Fisica ``E.R.
Caianiello'' and CNISM,
Universit\`a di Salerno, Via Allende 1, 84081 Baronissi (SA), Italy}

\date{\today}

\begin{abstract}
We study an atomic Josephson junction (AJJ) in presence of two
interacting Bose-Einstein condensates (BECs) confined in a double
well trap. We assume that bosons of different species interact
with each other. The macroscopic wave functions of the two
components obey to a system of two 3D coupled Gross-Pitaevskii
equations (GPE). We write the Lagrangian of the system, and from
this we derive a system of coupled ordinary differential equations
(ODE), for which the coupled pendula represent the mechanic
analogous. These differential equations control the dynamical
behavior of the fractional imbalance and of the relative phase of
each bosonic component.  We perform the stability analysis around
the points which preserve the symmetry and get an analytical
formula for the oscillation frequency around the stable points.
Such a formula could be used as an indirect measure of the
inter-species $s$-wave scattering length.  We also study the
oscillations of each fractional imbalance around zero and non zero
- the macroscopic quantum self-trapping (MQST) - time averaged
values. For different values of the inter-species
interaction amplitude, we carry out this study both by directly solving the two GPE and by solving the
corresponding coupled pendula equations. We show that, under certain conditions,
the predictions of these two approaches are in good agreement.
Moreover, we calculate the crossover value of the inter-species
interaction amplitude which signs the onset of MQST.
\end{abstract}

\submitto{\JPB}
\maketitle

\section{Introduction}

The prediction \cite{einstein} of Bose-Einstein condensation and
the realization in the laboratory of BECs \cite{anderson} paved
the way to a lot of  important theoretical and experimental
developments. One of these is the study of the atomic counterpart
\cite{leggett,smerzi,salasnich,salerno} of the Josephson effect
which occurs in superconductor-oxide-superconductor junctions
\cite{book-barone}. In
\cite{leggett,smerzi,salasnich,salerno}, realization of AJJ are
taken into account from a theoretical point of view. Few years
ago, Albiez {\it et al.} \cite{albiez} provided an experimental
realization of the AJJ. In 2007 Gati {\it et al.} \cite{gati} reviewed the experimental
realization of the AJJ focusing on the  data obtained experimentally with the
predictions of a many-body two-mode model \cite{milburn} and a mean-field description. Under certain conditions, a coherent
transfer of matter consisting of condensate bosons flows across
the junction. In the above references the AJJ physics is explored
in presence of a single bosonic species.  The possibility to
manage via magnetic and optical Feshbach resonances the intra- and
inter-species interactions \cite{minardi,papp} makes BECs mixtures
very promising candidates to successfully investigate  quantum
coherence and nonlinear phenomena such as the existence of
self-trapped modes and intrinsically localized states. Localized
states induced by the nonlinearity were shown to be quite generic
for multicomponent systems in external trapping potentials. In
particular, the emergence of coupled bright solitons from the
modulational instability of binary mixtures of BECs in optical
lattices was found numerically in \cite{kostov04}. More
sophisticated coupled localized states of two-component
condensates both in optical lattices and in parabolic traps were
reported in \cite{malomed04}. The existence of dark-bright states
of binary BECs mixtures was demonstrated in \cite{kivshar04}. On
the other hand, the existence of localized states of different
symmetry type (mixed symmetry states)  was numerically and
analytically demonstrated in \cite{Cruz07}. Properties of coupled
gap solitons in binary BECs mixtures with repulsive interactions
were also analyzed in the multidimensional case  \cite{malomed} as
well as for combined linear and nonlinear optical lattices
\cite{abdullaev08}. Although gap-soliton breathers of
multicomponent GPE involving periodic oscillations of the two
components densities localized on adjacent sites of an optical
lattice have been found \cite{Cruz07} (in analogy to what was done
for single component case in \cite{salerno}, such states can also
be seen as matter wave realizations of Josephson junctions), no
much numerical and theoretical study has been done until now on
AJJ of binary mixtures.

Recently, this has been considered in \cite{xu,satja} for the case
of a bosonic binary mixtures trapped in a double well potential,
for which a coupled pendula system of ODE for the temporal
evolution of the relative population and relative phase of each
component, was derived. Using this reduced system, the authors of
\cite{satja} have predicted the analogous of the macroscopic
quantum self-trapping phenomenon for a single bosonic component
\cite{smerzi}. No comparison between the reduced ODE system and
the full GPE dynamics has been performed, so that the question of
the validity of such prediction remains open. For single component
condensates, Salasnich {\it et al.} \cite{salasnich} have shown
that  a good agreement exists between the results obtained from
the GPE and those of the ODE. Similar agreement was obtained in
\cite{salerno} for AJJ realized with weakly interacting solitons
localized in two adjacent wells of an optical lattice. However,
the situation may be quite different for multicomponent
condensates, due to the interplay of intra- and inter-species
interactions which enlarge the number of achievable states (for
istance, mixed symmetry states can exist only in presence of the
inter-species interaction) as well as their stability, making the
system much more complicated.

The aim of the present paper is just to perform a systematic
investigation of possible  Josephson oscillations which can arise
in binary BECs mixtures trapped in a double well potential, as a
function of the system parameters. To this regard, we derive the
reduced coupled pendula system proceeding from a Lagrangian
formulation and from the canonical equations of motion. We show
that for certain conditions and range of parameters there exists a
good agreement between the solutions of the two GPE and the
predictions provided by the coupled pendula equations. We look for
the stationary points that preserve the symmetry and study their
stability; we get an analytical formula for the oscillation
frequencies around the equilibrium points. This formula shows the
possibility to determine the inter-species $s$-wave scattering
length from the frequency.

We analyze the influence of the inter-species interaction on the
temporal evolution of each relative population. In particular, by
employing the coupled pendula equations we show the existence of
MQST when the inter-species interaction amplitude is greater than
a certain value, for which we are able to provide an analytical
formula. As done by Satjia {\it et al.} \cite{satja}, we calculate
the values of the relative populations associated to the degenerate GPE states
that break the symmetry of the fractional imbalances. In addition, we perform the stability analysis by
explicitly calculating the associated oscillation frequencies. We,
moreover, show that the MQST-like evolution obtained by solving
the coupled pendula equations is close to that one obtained
by integrating the two coupled GPE.

Proceeding from the works of Albiez {\it et al.}
\cite{albiez} and of Gati {\it et al.} \cite{gati}, we correlate
our theoretical work with  experiments. Finally, we draw our
conclusions.

\section{AJJ with two bosonic species: quasi-analytical approach}

We consider two Bose-Einstein condensates of repulsively interacting Bosons
with different atomic species denoted below by $1$ and $2$. We
suppose that the two BECs are confined in a double-well trap
produced, for example, by a far off-resonance laser barrier that
separates each trapped condensate in two parts, L (left) and R
(right). We assume, moreover, that the two condensates interact
with each other.  In the mean field approximation, the macroscopic wave functions
$\Psi_i({\bf r},t)$, ($i=1,2$), of the interacting Bose-Einstein condensates
in a trapping potential $V_{trap}({\bf r})$ at zero-temperature
satisfy the two coupled Gross-Pitaevskii equations
\beq \label{GPE} i \hbar \frac{\partial \Psi_i}{\partial t} =
-\frac{\hbar^2}{2m_i}\nabla^2 \Psi_i+[V_{trap}({\bf r})
+g_{i}|\Psi_i|^2+g_{ij}|\Psi_j|^2]\Psi_i. \eeq
Here $\nabla^2$ denotes the 3D Laplacian, and $\Psi_i({\bf r},t)$
is subject to the normalization condition 
\beq
\label{normalizationwf} \int \int \int dxdydz \,
|\Psi_i(x,y,z)|^2=N_i \; ,\eeq with $N_i$ the number of bosons of
the $i$th species. Similarly, $m_i$ , $a_i$ and $g_{i}=4\pi \hbar^2 a_i/m_i$  denotes the atomic
mass, the $s$-wave scattering length  and the
intra-species coupling constant of the $i$th species; $g_{ij}=2\pi \hbar^2 a_{ij}/m_r$ ($i
\neq j$) is the inter-species coupling constant, with $m_r=m_i
m_j/(m_i+m_j)$ the reduced mass, and $a_{ij}$ the associated
$s$-wave scattering length.
\begin{figure}[h]
\centerline{
\includegraphics[width=10cm,clip]{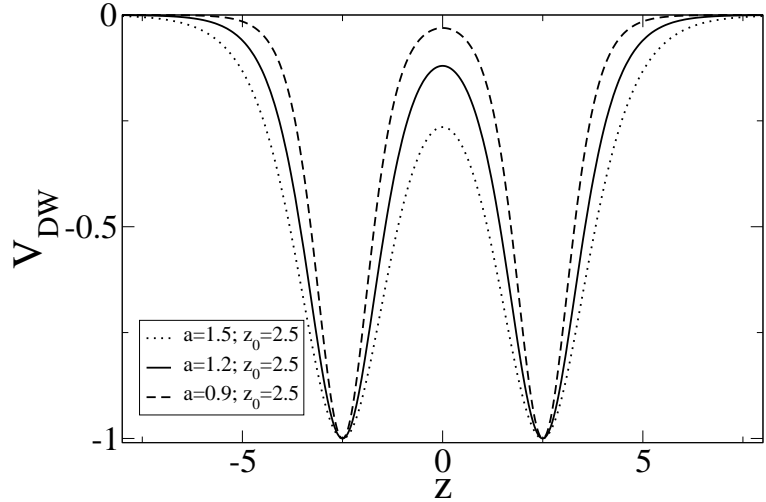}}
\caption{The double well potential (\ref{PT})  as a function of
$z$. Lengths are measured in units of $a_{\bot,1}=a_{\bot,2}
\equiv a_{\bot}$, and energies in units of $\hbar \omega_1=\hbar
\omega_2 \equiv \hbar \omega$.} \label{fig1}
\end{figure}
In the following, we shall consider both $g_i$ and $g_ {ij}$ as
free parameters, due to the possibility to change the scattering
lengths $a_i$ and $a_{ij}$ at will by using the technique of Feshbach
resonances. Here we take into account the case in which the two
BECs interact attractively, see \cite{minardi} and \cite{simoni}.

The trapping potential for both components is taken to be the
superposition of a strong harmonic confinement in the radial
($x$-$y$) plane and of a double well (DW) potential in the axial
($z$) direction. For the $i$th component, we model this trapping
potential in the form
\beq \label{trap} V_{trap}({\bf
r})=\frac{m_i \omega_i^2}{2} (x^2+y^2)+V_{DW}(z)\; ,\eeq where, for
symmetric configurations in the $z$ direction, we model the DW
potential as
\begin{eqnarray}
\label{PT} &&V_{DW}(z)=V_{L}(z)+V_{R}(z),\;\;
V_{L}(z)=-V_{0}\bigg[Sech^2(\frac{z +z_0}{a})\bigg],\nonumber\\
&&V_{R}(z)=-V_{0}\bigg[Sech^2(\frac{z -z_0}{a})\bigg],\;\;
V_{0}=\hbar \omega_i\big[1+Sech^2(\frac{2z_0}{a})\big]^{-1} ,\nonumber\\
\;\end{eqnarray} i.e. the combination of two P\"{o}schl-Teller (PT)
potentials, $V_L(z)$ and $V_R(z)$, separated by a potential barrier, the height of which can be changed by varying
$a$, centered around the points $-z_0$ and $z_0$ (see Fig. 1). Note
that the usage of PT potentials is only for the benefit of
improving accuracy in our numerical GPE calculations (see below),
taking advantage of the integrability of the underlying linear
system. We remark, however, that the obtained results are of
generic validity also for more confining (e.g. not saturating to
zero at large distances) double well potentials.

We are interested to study the dynamical oscillations of the populations of each
condensate between the left (L) and right (R) wells when the the barrier is large enough so that the link is weak.
To exploit the strong harmonic confinement  in the ($x$-$y$) plane and  get the effective 1D equations describing the dynamics in the $z$ direction,
we write the Lagrangian associated to the GPE equations
in (\ref{GPE})
\begin{eqnarray}
\label{lagrangian} L &=& \int d^3 {\bf
r}\,\bigg(\bigg[\sum_{i=1,2} \bar \Psi_i\big(i \hbar
\frac{\partial}{\partial t}+\frac{\hbar^2}{2m_i} \nabla^2
\big)\Psi_i\nonumber\\ &-&V_{trap}({\bf
r})|\Psi_i|^2-\frac{g_{i}}{2}|\Psi_i|^4\bigg]-g_{ij}|\Psi_i|^2|\Psi_j|^2\bigg)\;,\end{eqnarray}
where $\bar \Psi_i$ denotes the complex conjugate
of $\Psi_i$, and $i \neq j$ and adopt the ansatz \cite{salasnichvecsol}
\beq
\label{ansatz} \Psi_i(x,y,z,t)=\frac{1}{\sqrt{\pi} a_{\bot,i}}
\exp \bigg[-\frac{x^2+y^2}{2 a_{\bot, i}^2}\bigg] f_{i}(z,t)\;,
\eeq
where $\displaystyle{a_{\bot,i}=\sqrt{\frac{\hbar}{m_i
\omega_i}}}$ and  the $f_i(z,t)$ obey
$\int_{-\infty}^{+\infty}dz|f_i(z)|^2=N_i$, so that the normalization condition
Eq. (\ref{normalizationwf}) is satisfied. By inserting this ansatz (\ref{ansatz})
in (\ref{lagrangian}) and performing the integration in the radial
plane, we obtain the effective 1D Lagrangian for  the fields $f_i(z,t)$
\begin{eqnarray}
\label{effectivelagrangian} \tilde L &=& \int d
z\,\bigg(\bigg[\sum_{i=1,2}
\bar f_i\big(i \hbar \frac{\partial}{\partial
t}+\frac{\hbar^2}{2m_i}\frac{\partial^2}{\partial z^2} \big)f_i \nonumber\\
&-& (\epsilon_i+V_{DW}(z))|f_i|^2-
\frac{\tilde g_{i}}{2}|f_i|^4\bigg]- \tilde g_{ij}|f_i|^2|f_j|^2\bigg)\;,\nonumber\\
\end{eqnarray}
where the effective parameters for the 1D dynamics are given in terms of the original ones by:
$\displaystyle{\epsilon_i=\frac{\hbar^2}{2m_i
a_{i,\bot}^2}+\frac{m_i \omega_i^2 a_{i,\bot}^2}{2}} $,
$\displaystyle{\tilde g_{i}=\frac{g_{i}}{2 \pi a_{\bot,i}^2}}$,
and $\displaystyle{\tilde g_{ij}=\frac{g_{ij}}{\pi
(a_{\bot,i}^2+a_{\bot,j}^2)}}$.
By varying $\tilde L$ with respect
to $\bar f_i$, we obtain the 1D GPE for the field $f_i$
\begin{equation}
\label{1dGPE}  i \hbar \frac{\partial f_i}{\partial t} =
-\frac{\hbar^2}{2m_i}\frac{\partial^2 f_i}{\partial
z^2}+[\epsilon_i+V_{DW}(z) + \tilde g_{i}|f_i|^2+ \tilde
g_{ij}|f_j|^2]f_i \; .\end{equation}

It is possible to study
the AJJ dynamics described by Eq. (\ref{1dGPE}) by using the
two-mode approximation discussed by Milburn {\it et al.} in \cite{milburn}. In particular,
we assume, for each $f_i$, the following time-dependent wave
function decomposition
\begin{eqnarray}
\label{variational} f_i(z,t) =
\psi^{L}_i(t)\phi^{L}_i(z)+\psi^{R}_i(t)\phi^{R}_i(z) \;
,\end{eqnarray} where \beq \label{guess}
\psi^{\alpha}_i(t)=\sqrt{N^{\alpha}_i} e^{i
\theta^{\alpha}_i(t)}\; ,\eeq with $\alpha$=L,R, and a constant
total number of particles given by
$N^L_i+N^R_i=|\psi^{L}_i(t)|^2+|\psi^{R}_i(t)|^2 \equiv N_i$, with
$\int_{-\infty}^{+\infty} dz|\phi^{\alpha}_i(z)|^2 =1$ and
$\int_{-\infty}^{+\infty} dz\phi^{L}_i(z) \phi^{R}_i(z) =0$.
Neglecting terms of order greater than two in the overlaps of the
$\phi$'s, we can  write the Lagrangian (\ref{effectivelagrangian})
in terms of $N^{\alpha}_i$ and $\theta^{\alpha}_i$ as

\begin{eqnarray}
\label{lagrangianbis} \bar L &=&
\sum_{i=1,2}\bigg[-\hbar\dot{\theta}^{L}_iN^{L}_i-\hbar\dot{\theta}^{R}_iN^{R}_i
-E^{L}_iN^{L}_i-E^{R}_iN^{R}_i\nonumber \\ &+& 2 K_i \sqrt{N^{L}_i
N^{R}_i} \cos(\theta^{L}_i-\theta^{R}_i)
-\big(\frac{U^{L}_i}{2}(N^{L}_i)^{2}+\frac{U^{R}_i}{2}(N^{R}_i)^{2}\big)\bigg]\nonumber\\
&-&U_{12}^{L}N^{L}_1N^{L}_2-U_{12}^{R}N^{R}_1N^{R}_2\,,\end{eqnarray}
where
\begin{eqnarray}
\label{parameters} && E^{\alpha}_i = \int d z
\,\bigg[\frac{\hbar^2}{2m_i}\,(\frac{d \phi^{\alpha}_i} {d z})^2
+\big(V_{DW}+\frac{\hbar^2}{2m_i
a_{\bot,i}^2}+\frac{m_i\omega_i^2a_{\bot,i}^2}{2}\big)(\phi^{\alpha}_i)^2\bigg],
\nonumber\\
&&U^{\alpha}_{i}=\tilde g_{i} \int dz \,
(\phi^{\alpha}_i)^4,\;\;\;\; U^{\alpha}_{12}=\tilde g_{12} \int dz
\, (\phi^{\alpha}_1)^2
(\phi^{\alpha}_2)^2,\nonumber\\
&&K_i=-\int dz
\,\bigg[\frac{\hbar^2}{2m_i}\frac{d \phi^{L}_{i}}{d
z}\frac{d \phi^{R}_i}{d z}
+ V_{DW}\phi^{L}_i\phi^{R}_i\bigg]\; .\nonumber\\
\end{eqnarray}

One may get a good approximation for the functions $\phi^{L}_{i}(z)$ and $\phi^{R}_{i}(z)$
when the double well potential  $V_{DW}(z)$ is such that the two lowest energy
eigenvalues of the corresponding Schr{\"{o}}dinger equation
constitute a closely spaced doublet well separated  from the higher excited
levels, and the $\tilde g$'s are not too large (see, for example, \cite{milburn}).
If the real symmetric function $\phi^{S}_i(z)$ and the real
antisymmetric function $\phi^{A}_i(z)$ are the wave functions of
the ground state and of the first excited state, respectively, then  $\phi^{L}_{i}(z)$ and $\phi^{R}_{i}(z)$  may be chosen as
\begin{equation}
\label{firstmethod}
\phi^{L}_{i}(z)=\frac{\phi^{S}_i(z)+\phi^{A}_i(z)}{\sqrt{2}}\;,\;\;\;\;\;
\phi^{R}_{i}(z)=\frac{\phi^{S}_i(z)-\phi^{A}_i(z)}{\sqrt{2}}\;.
\end{equation}

Remember that $\phi^{S}_i(z)$ and
$\phi^{A}_i(z)$ satisfy the
relations: $\int_{-\infty}^{+\infty} dz|\phi^{S}_i(z)|^2
=\int_{-\infty}^{+\infty} dz|\phi^{A}_i(z)|^2 =1$ and
$\int_{-\infty}^{+\infty} dz \phi^{A}_i(z) \phi^{S}_i(z) =0$.
Having chosen $V_{DW}(z)$ as the sum of two of two P\"{o}schl-Teller (PT) wells
(see Eq. (\ref{PT})), the
functions $\phi^{L}_{i}(z)$ and $\phi^{R}_{i}(z)$ may be
analytically calculated following a perturbative approach. Let us
consider the eigenvalues problem corresponding to Eq.
(\ref{1dGPE}) with $\tilde g_i= \tilde g_{ij}=0$. We know exactly
the wave functions for this eigenvalues problem when the potential
is given by a single $V_{\alpha}(z)$ ($\alpha=L,R$), for example
$V_{L}(z)$. The wave function of the ground state is \cite{landau}
\begin{eqnarray}
\label{wfpt} &&\phi_{i}^{(L,PT)}(z)=A\big[1-Tanh^{2}(\frac{z +
z_0}{a})\big]^{B_{i}/2} \nonumber\\
&& B_{i}=-\frac{1}{2}+\sqrt{\frac{2m_{i}V_0
a^2}{\hbar^{2}}+\frac{1}{4}} \; .\end{eqnarray} In Eq.
(\ref{wfpt}) $A$, equal for both sides, ensures the normalization
of the wave function. Since we are assuming that the two lowest
energetic levels are well separated from the higher ones, when the
potential is perturbed by the presence of $V_{R}(z)$, we look for
the eigenstates in the form of a linear superposition of
$\phi_{i}^{(L,PT)}$ and $\phi_{i}^{(R,PT)}$. For each component,
to the first order of such a perturbative theory, the ground state
wave function  $\phi_{i}^{S}(z)$ and the first excited state wave
function $\phi_{i}^{A}(z)$ read
\begin{eqnarray}
\label{pt}
&&\phi_{i}^{S}(z)=M_{S}\big(\phi_{i}^{(L,PT)}(z)+\phi_{i}^{(R,PT)}(z)\big) \nonumber\\
&&\phi_{i}^{A}(z)=M_{A}\big(\phi_{i}^{(L,PT)}(z)-\phi_{i}^{(R,PT)}(z)\big)
\; .\end{eqnarray}  Here $M_{S}$ and $M_{A}$ ensure the
normalization of $\phi_{i}^{S}(z)$ and $\phi_{i}^{A}(z)$,
respectively. Note that $K_i$ is equal to $(E^{A}_i-E^{S}_i)/2$, with $E^{S}_i$ and
$E^{A}_i$, also perturbatively calculated, the
energies associated to $\phi^{S}_i(z)$ and $\phi^{A}_i(z)$,
respectively. The quantity
\beq \label{fr}
\omega=\frac{(E^{A}_i-E^{S}_i)}{\hbar}
 \eeq
is the Rabi frequency. This frequency characterizes the oscillations of a
particle between the states $\phi^{L}_{i}$ and $\phi^{R}_{i}$. By using Eq.
(\ref{pt}) in the decomposition (\ref{firstmethod}), we are able to write the
functions $\phi^{L}_{i}(z)$ and $\phi^{R}_{i}(z)$ in terms of $\phi_{i}^{(L,PT)}(z)$
and $\phi_{i}^{(R,PT)}(z)$ in the following way
\begin{eqnarray}
\label{finaldecomposition}
&& \phi^{L}_{i}=\frac{\bigg[(M_S+M_A)\phi_{i}^{(L,PT)}+(M_S-M_A)\phi_{i}^{(R,PT)}\bigg]}{\sqrt{2}} \nonumber\\
&&
\phi^{R}_{i}=\frac{\bigg[(M_S-M_A)\phi_{i}^{(L,PT)}+(M_S+M_A)\phi_{i}^{(R,PT)}\bigg]}{\sqrt{2}}
\;. \nonumber\\\end{eqnarray} \\
Note that $\phi^{S}_i(z)$ and $\phi^{A}_i(z)$, and the associated energies, may be
numerically found as the wave functions of the two lowest states of the eigenvalues problem corresponding to Eq.
(\ref{1dGPE}) in absence of interactions. Then, by using the decomposition
(\ref{firstmethod}), one calculates the functions $\phi^{L}_i(z)$ and
$\phi^{R}_i(z)$.
We have verified that the perturbative theory provides practically the same results as the
numerical approach .\\

Let us, now, focus on the Lagrangian (\ref{lagrangianbis}). The
conjugate moments of the generalized coordinates $N^{\alpha}_i$
and $\hbar\theta^{\alpha}_i$ are given by
\begin{equation}
\label{conjugatemoments} p_{N^{\alpha}_i}=\frac{\partial \bar
L}{\partial \dot{N}^{\alpha}_i}=0\;,\;\;\;\;\;
p_{\theta^{\alpha}_i}=\frac{1}{\hbar}\frac{\partial \bar
L}{\partial \dot{\theta}^{\alpha}_i}=-N^{\alpha}_i\; .
\end{equation}
The Hamiltonian of the system is
\begin{eqnarray}
\label{hamiltonian} H &=&
-\sum_{i=1,2}[p_{\theta_{i}^{L}}E_i^{L}+p_{\theta_{i}^{R}}E_i^{R}]
- \sum_{i=1,2}2K_i\sqrt{p_{\theta^{L}_i}p_{\theta^{R}_i}}\cos(\theta^{L}_{i}-\theta^{R}_{i})+\nonumber\\
&+&\sum_{i=1,2}[\frac{U_{i}}{2}^{L}p_{\theta^{L}_i}^2+\frac{U_{i}^{R}}{2}p_{\theta^{R}_i}^2]
+U_{12}^{L}p_{\theta^{L}_{1}} p_{\theta^{L}_{2}} +
U_{12}^{R}p_{\theta^{R}_1} p_{\theta^{R}_2}  \; .
\end{eqnarray}

The evolution equations for the fractional
imbalance $z_i=(N^{L}_i-N^{R}_i)/N_i$ and for the relative phase
$\theta_i=\theta^{R}_i-\theta^{L}_i$ for each component are
derived from the canonical equations associated to the Hamiltonian
(\ref{hamiltonian})
\begin{equation}
\label{canonical}
\dot{p}_{\theta^{\alpha}_{i}}=-\frac{1}{\hbar}\frac{\partial
H}{\partial \theta^{\alpha}_{i}}\;, \;\;\;\;\;
\dot{\theta}^{\alpha}_{i}=\frac{1}{\hbar} \frac{\partial
H}{\partial p_{\theta^{\alpha}_{i}}} \; .
\end{equation}
By subtracting the equation for $\dot{p}_{\theta^{i}_{L}}$ from
the one for $\dot{p}_{\theta^{i}_{R}}$, we obtain the equation for
the temporal evolution of $z_{i}$. Similar arguments leads to the
equation for $\theta_i(t)$. In the following, we shall assume two
wells to be symmetric, i.e. $E_{i}^{L}=E_{i}^{R}$,
$U_{i}^{L}=U_{i}^{R} \equiv U_{i}$, $U_{12}^{L}=U_{12}^{R} \equiv
U_{12}$. The fractional imbalance and the relative phase for each
component vary in  time according to the following (coupled
pendula) equations:
\begin{eqnarray}
\label{evolution} &&\dot{z_i}=-\frac{2}{\hbar}K_i\sin \theta_i
\sqrt{1-z_i^2}\nonumber\\
&&\dot{\theta_i}=\frac{U_{i}N_iz_i}{\hbar} +\frac{2K_i
z_i}{\hbar\sqrt{1-z_i^2}}\cos \theta_i+\frac{U_{12} N_j
z_j}{\hbar}
\; .\nonumber\\
\end{eqnarray}
Note that when $U_{12}=0$,  equations (\ref{evolution}) reduce to
the usual equations for a single component obtained in
\cite{smerzi}. At this point, we observe that it is possible to obtain
Eq. (\ref{evolution}) proceeding from the following equations
\begin{equation}
\label{canonical2} N_i\dot{z_i}=-\frac{1}{\hbar}\frac{\partial
\tilde H}{\partial \theta_i}\;,\;\;\;\;\; N_i
\dot{\theta_i}=\frac{1}{\hbar} \frac{\partial \tilde H}{\partial
z_i} \; ,\end{equation} where $\tilde H$ is
\begin{equation}
\label{hamiltonianbis} \tilde H = -\sum_{i=1,2}2K_i
N_i[\sqrt{(1-z_i^2)}\cos \theta_i]
+\sum_{i=1,2}\frac{U_{i}}{2}N_i^2 z_i^2 + U_{12} N_1 N_2 z_1z_2
.\end{equation} To fix ideas, let us consider, as done in
\cite{papp}, a mixture of two bosonic isotopes of the same atom,
so that one may have different intra-species and inter-species
interactions. For simplicity, for the time being we will neglect
the mass difference between the two species. We compare the
temporal behavior of the fractional imbalances obtained
integrating Eq. (\ref{1dGPE}) with the $z_i(t)$ obtained by
solving the coupled differential equations (\ref{evolution}). To
this end, in the wave functions (\ref{wfpt}) we assume that the
quantities $B_i$ are the same for both species, and in the double
well potential (\ref{PT}) we set $a=1.2$ and $z_0=2.5$, as illustrated by the continuous line of Fig. 1. We calculate
the parameters $U_{i}$, $U_{12}$, and $K_1=K_2 \equiv K$ by using
 Eqs. (\ref{parameters}). Note that the value of $K$ (equal,
for the above values of $a$ and $z_0$, to $0.0148$), provided by
the last formula of Eqs. (\ref{parameters}) coincides with
$\displaystyle{\frac{\hbar \pi}{\tau_0}}$. Here $\tau_0$ is the
oscillation period of $z_i(t)$ obtained by numerically solving the
1D GPE (\ref{1dGPE}) when $U_{i}=U_{12}=0$. This period is just
equal to $2\pi/\omega$ with $\omega$ given by Eq. (\ref{fr}). In
Fig. 2, the first panel of each $z_{i}(t)$ graph shows the perfect
agreement between the two coupled 1D GPE and the coupled pendula
equations when the bosons do not interact at all. In the second
panel of each $z_{i}(t)$ of Fig. 2, $U_{i}$ is finite and
$U_{12}=0$. Finally, in the last panel, $U_{i}$ and $U_{12}$ are
both finite. From the second and the third panels of Fig. 2, we
see that the solutions of the ODE system (\ref{evolution}) with
$K=0.0148$ (dot-dashed lines) shows a certain displacement with
respect to the ones (dashed lines) predicted by solving the two 1D
GPE (\ref{1dGPE}). The continuous lines, in the second and in the
third panels of Fig. 2, represent the solutions of the ODE system
(\ref{evolution}) with $K$ obtained via a fitting procedure.
The best-fit $K$'s that we found are equal to $0.0155$ and $0.0151$
when only $U_{i} \neq 0$ and when $U_{i}$ and $U_{12}$ are both
finite, respectively.
Notice that these values are very close to the above theoretical estimate of $K$,
so the coupled pendula equations may be used to consistently describe the AJJ physics.\\
\begin{figure}[h]
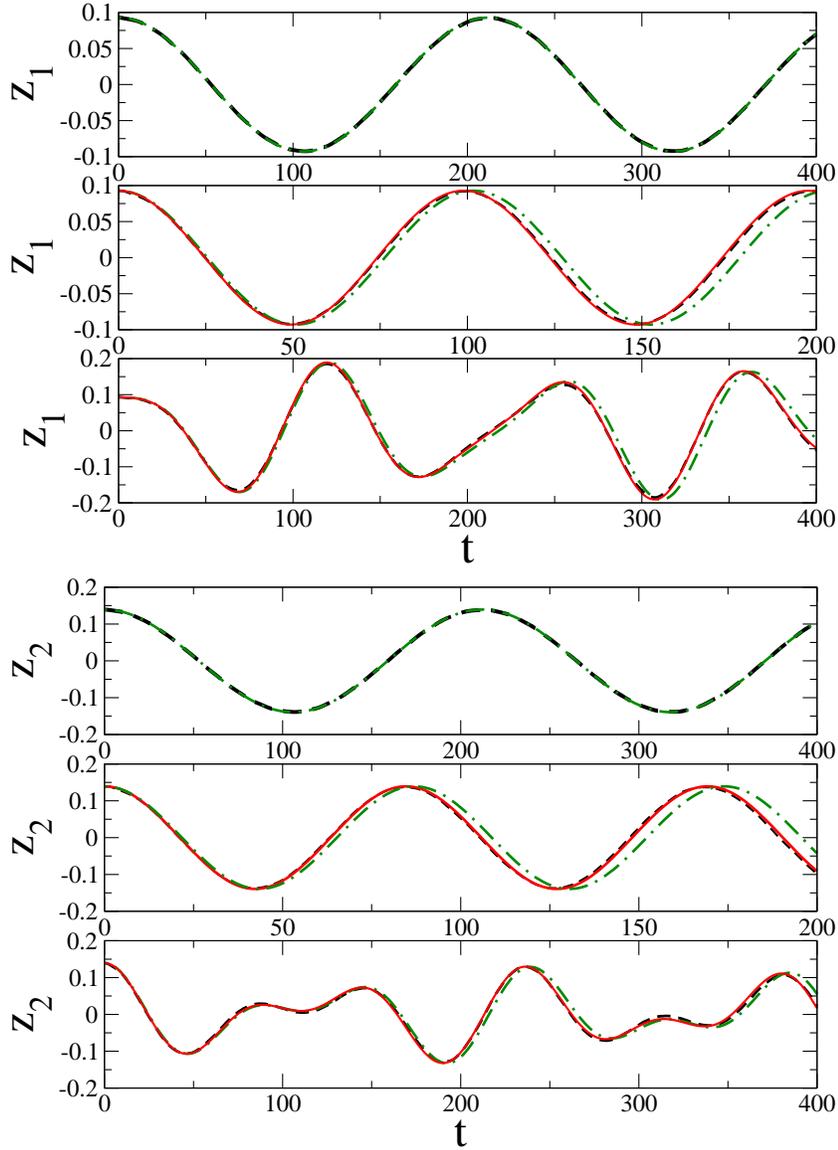

\centering
\begin{tabular}{cc}
\epsfig{file=p1.eps,width=0.7\linewidth,clip=}\\
\epsfig{file=p2.eps,width=0.7\linewidth,clip=}
\end{tabular}
\caption{Fractional imbalance of the two bosonic species vs. time.
Here $N_1=100$ and $N_2=150$. In each plot of $z_i(t)$, from top
to bottom: $U_{1}=U_{2}=U_{12}=0$; $U_{1}=U_{2} =0.001, U_{12}=0$;
$U_{1}=U_{2} \equiv U=0.001, U_{12}=-U/2$. The dashed line
represents data from integration of Eq. (\ref{1dGPE}), the
dot-dashed line represents solution of Eq. (\ref{evolution}) with
$K_1=K_2 \equiv K=0.0148$, and the continuous line, in the second and third
panels, represents solution of Eq. (\ref{evolution}) with the best-fit
$K$'s, say $K_{bf}$. For each $z_i(t)$, $K_{bf}=0.0155$
($U_{1}=U_{2} =0.001, U_{12}=0$) and $K_{bf}=0.0151$ ($U_{1}=U_{2}
\equiv U=0.001, U_{12}=-U/2$). We used the initial conditions
$z_1(0)=0.1$, $z_2(0)=0.15$, and $\theta_i(0)=0$. Time is measured
in units of $(\omega_1)^{-1}=(\omega_2)^{-1} \equiv \omega^{-1}$,
lengths are measured in units of $a_{\bot,1}=a_{\bot,2} \equiv
a_{\bot}$, and energies in units of $\hbar \omega$.} \label{fig2}
\end{figure}

To gain physical insight in the behavior of the system, we carry
out the stability analysis. We study the oscillations around the
points for which the time derivatives of $z_i$ and of $\theta_i$
are zero, i.e. around the stationary points. We diagonalize the
Jacobian matrix associated to Eqs. (\ref{evolution}). The
eigenvalues $\lambda$ allows us to determine the frequencies of
oscillation around the points of equilibrium. If the $\lambda$'s
are of the form $i \omega$ (with $\omega$ a pure real number),
$\omega$ will be the oscillation frequency around stable
equilibrium points. By performing the above analysis, we obtain
the following classes of stationary points:
\begin{enumerate}
\centering
\item $z_1=0\;,z_2=0\;, \theta_1=0\;, \theta_2=0$\;\;\;
\item $z_1=0\;, z_2=0\;, \theta_1=0, \theta_2=\mp \pi$
\item  $z_1=0, z_2=0, \theta_1=\mp \pi, \theta_2=0$
\;\; \item $z_1=0, z_2=0, \theta_1=\mp \pi, \theta_2=\mp \pi$,\\
$z_1=0, z_2=0, \theta_1=\mp \pi, \theta_2=\pm \pi$ .
\end{enumerate}
We observe that the points of a given class are characterized by
the same $\lambda$. The stationary solution (i) represents a
stable equilibrium if $U_{12}$ satisfies the condition
\begin{eqnarray} && \label{conditioni}
-\Lambda^{(1)}< U_{12}<\Lambda^{(1)} \nonumber\\
&&
\Lambda^{(1)}=\frac{\sqrt{(2K_1+N_1U_{1})(2K_2+N_2U_{2})}}{\sqrt{N_1N_2}}\;
\end{eqnarray}\\
provided
\beq \label{providedi} U_{i}>-\frac{2K_i}{N_i} \; .\eeq
The small amplitude oscillations frequency, say $\omega^{(1)}$,
around the point (i) is
\begin{eqnarray}
\label{f1} \omega^{(1)}_{\pm}&=&\frac{1}{\hbar}\big(K_1(2K_1+U_{1}N_1)
+ K_2( 2K_2+ U_{2}N_2)  \pm  \Delta\big)^{1/2}\;, \nonumber\\
\Delta &=& \bigg[-4 K_1 K_2 (4 K_1 K_2+2 K_1 U_{2} N_2 + 2K_2
U_{1}N_1- U_{12}^2N_1N_2 \nonumber \\ &+& U_{1} U_{2}N_1N_2) +
\big(K_1(2K_1+U_{1}N_1)+ K_2(2K_2+ U_{2}N_2)\big)^2\bigg]^{1/2}\;
\end{eqnarray}
with $+$ and $-$ corresponding to the normal modes of the
linearized system associated to Eq. (\ref{evolution}). When
$z_1(0)=\mp z_2(0) \ll 1$, $U_{1}=U_{2}$, $K_1=K_2$, and
$N_1=N_2$, the fractional imbalances $z_i$ oscillate around the
point (i) according to the law 
\beq \label{coslawi}
z_{i}(t)=z_{i}(0) \cos \omega^{(1)}_{\pm} t  \: .\eeq
Let us operate  in Eqs.
(\ref{conditioni}), (\ref{providedi}), and (\ref{f1}) the
replacement $U_{i} \rightarrow -U_{i}$. Then, we obtain the
stability conditions and the oscillation frequency associated to
the class (iv). We have verified that for the stationary points of
type (ii), under certain conditions (analytically achievable but
very complicated), the eigenvalues of Jacobian matrix are all of
the form $i \omega$. The small amplitude oscillations frequency,
say $\omega^{(2)}$,  around the point (ii) is
\begin{eqnarray}
\label{f2} \omega^{(2)}_{\pm}&=&\frac{1}{\hbar}\big(K_1(2K_1+U_{1}N_1)
+ K_2( 2K_2- U_{2}N_2) \pm \Delta\big)^{1/2}\;, \nonumber\\
\Delta &=&\bigg[-4 K_1 K_2 (4 K_1 K_2-2 K_1 U_{2} N_2 +2K_2
U_{1}N_1+ U_{12}^2N_1N_2\nonumber\\&-& U_{1} U_{2}N_1N_2)
+\big(K_1(2K_1+U_{1}N_1)+ K_2(2K_2- U_{2}N_2)\big)^2\bigg]^{1/2}\;
.\end{eqnarray} 
For the oscillations of $z_i$ around the point (ii), one may use
arguments analogous to the ones employed for the class (i). 
If we start, now, from the points of the class (ii), and replace $U_{i}$
with $-U_{i}$, we get the stability regions and the oscillation
frequency for the point of type (iii). Let us focus, to fix the ideas, on the
frequency (\ref{f1}), and on the formula of Eq.
(\ref{parameters}) which gives the inter-species interaction
amplitude $U_{12}$. We note that $\tilde g_{12}$ is directly related to the
inter-species $s$-wave scattering length. This quantity, then, can
be determined from the oscillation frequency (\ref{f1}) once one keeps fixed $K_{i}$, $U_{i}$, and $N_{i}$. We
will discuss this point with more details in the following.

At this point, it is worth observing that, because of the non
linearity associated to the inter- and intra-species interactions, there is a
class of degenerate GPE eigenstates that breaks the $z_i$
symmetry. Let assume that $U_{1}=U_{2} \equiv U$, $K_1=K_2 \equiv
K$, and $N_{1}=N_{2} \equiv N$. In correspondence of
$\theta_i=\pi$, we have looked for non zero stationary solutions
of the system (\ref{evolution}). We have found four classes of
fractional imbalances corresponding to the $z_i$ broken symmetry; we
have verified that two of these classes do not correspond to a
stable equilibrium. Let us consider the two classes describing
stable equilibrium, say $I$ and $II$. For the class $I$, we have
\begin{eqnarray}
\label{sbI}
&&z^{(I)}_{1,sb}=\pm \sqrt{1-\bigg(\frac{2K}{N(U + U_{12})}\bigg)^2}
\nonumber\\
&&z^{(I)}_{2,sb}=z^{(I)}_{1,sb}\;, \end{eqnarray} provided
$|(U+U_{12})|>2K/N$. When $0<U<2K/N$, the solution (\ref{sbI}) is always
stable, and the corresponding oscillation frequency is
\begin{eqnarray}
\omega_{A}^{(I)}=\frac{1}{\hbar}\sqrt{\bigg(N(U+U_{12})\bigg)^2-4 K^2}\;.
\end{eqnarray}
For $U>2K/N$, the solution (\ref{sbI}) is stable when
\begin{eqnarray}
&&U_{12}>\tilde U_{12}^{(I)}=\frac{2K}{N}\bigg(-UN/2K\nonumber\\
&+&\frac{3^{1/3}-(9UN/2K+\sqrt{3+81(UN/2K)^2})^{2/3}}{3^{2/3}
(9UN/2K+\sqrt{3+81(UN/2K)^2})^{1/3}}\bigg).\nonumber\\\;
\end{eqnarray}
The corresponding oscillation frequency is
\begin{eqnarray}
\omega_{B}^{(I)}=\frac{1}{\hbar}\sqrt{\bigg(N(U+U_{12})
\bigg)^2+4K^{2}\bigg(\frac{U_{12}-U}{U+U_{12}}\bigg)}
\; .
\end{eqnarray}

It is possible determine the crossover value, say
$U^{(I,cr)}_{12}$, of the inter-species interaction strength
signing the onset of the self-trapping. We start evaluating the Hamiltonian
(\ref{hamiltonianbis}) in $z_{i}=z^{(I)}_{i,sb}$ and
$\theta_i=\pi$. Let us denote the energy obtained in this way by $E^{(I)}$. This
energy reads
\beq \label{esbI} E^{(I)}=\frac{4
K^2}{U+U_{12}}+N^2(U+U_{12})\;.\eeq Then, we evaluate the
Hamiltonian (\ref{hamiltonianbis}) at $t=0$, i.e. $\tilde
H(z_i(0),\theta_i(0))$. We require that \beq \label{condition}
\tilde H(z_i(0), \theta_i(0))>4KN \; \eeq with $4KN$ the value got
by $E^{(I)}$ when $z^{(I)}_{i,sb}=0$, i.e. when
$2K=N(U+U_{12})$. Then, we get
\begin{eqnarray}
\label{u12critico}
U^{(I,cr)}_{12} &=& \frac{K}{N
z_1(0)z_2(0)}\bigg[4-\frac{UN}{2K}\sum_{i=1,2}z_i(0)^2 \nonumber\\
&+&2\sum_{i=1,2}\sqrt{1-z_i(0)^2}\cos\theta_i(0)\bigg]\;.\end{eqnarray}
When the condition $U_{12}>U^{(I,cr)}_{12}$ is satisfied, the system will be self-trapped.
For the solution of the type $II$, we have
\begin{eqnarray}
\label{sbII}
&&z^{(II)}_{1,sb}=\pm \sqrt{1-\bigg(\frac{2K}{N(U - U_{12})}\bigg)^2}\nonumber\\
&&z^{(II)}_{2,sb}=-z^{(II)}_{1,sb}\;, \end{eqnarray} provided that
$|(U-U_{12})|>2K/N$. When
$0<U<2K/N$, this solution is always stable and is
characterized by the oscillation frequency
\begin{eqnarray}
\omega_{A}^{(II)}=\frac{1}{\hbar}\sqrt{\bigg(N(U-U_{12})\bigg)^2-4 K^2}
\; .\end{eqnarray}
When $U>2K/N$, the solution (\ref{sbII}) is stable if
\begin{eqnarray}
&&U_{12}<\tilde
U_{12}^{(II)}=\frac{2K}{N}\bigg(UN/2K\nonumber\\
&-&\frac{3^{1/3}-(-9UN/2K+\sqrt{3+81(UN/2K)^2})^{2/3}}{3^{2/3}(-9UN/2K+
\sqrt{3+81(UN/2K)^2})^{1/3}}\bigg),\nonumber\\
\;
\end{eqnarray}
and the corresponding oscillation frequency reads
\begin{eqnarray}
\omega_{B}^{(II)}=\frac{1}{\hbar}\sqrt{\bigg(N(U-U_{12})\bigg)^2+
4K^2\bigg(\frac{U+U_{12}}{U_{12}-U}\bigg)} \; .\end{eqnarray}
Also for the solution $II$, it is possible to determine the
inter-species interaction amplitude which signs the self-trapping
onset, say $U^{(II,cr)}_{12}$, by using the same argument employed
for the class $I$. Also in this case, we proceed by evaluating the Hamiltonian
(\ref{hamiltonianbis}) in $z_{i}=z^{(II)}_{i,sb}$ and
$\theta_i=\pi$. Let us denote the energy obtained in this way by $E^{(II)}$.
This energy reads
\beq \label{esbII} E^{(II)}=\frac{4
K^2}{U-U_{12}}+N^{2}(U-U_{12})\;.\eeq
\begin{figure}
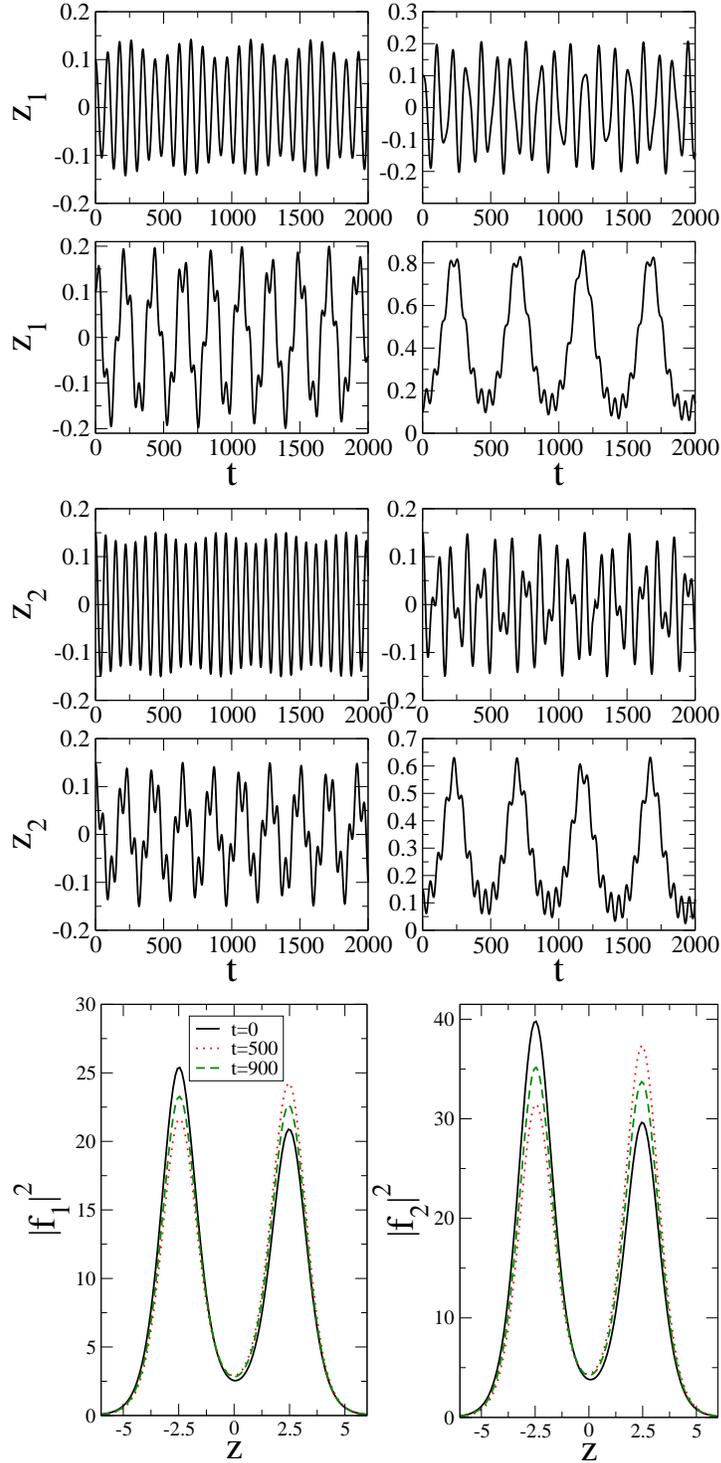

\centering
\begin{tabular}{cc}
\epsfig{file=z1revised.eps,width=0.6\linewidth,clip=}\\
\epsfig{file=z2revised.eps,width=0.6\linewidth,clip=} \\
\epsfig{file=f1f2revised.eps,width=0.6\linewidth,clip=}
\end{tabular}
\caption{Fractional imbalances of the first and second bosonic
species as functions vs. time. Here $N_1=100$ and $N_2=150$, $K_1=
K_2=0.0148$, and $U_{1}=U_{2} \equiv U=0.1 K_1$. Plots are for
different values of $U_{12}$. $U_{12}=-U/20$ and $U_{12}=-U/2$
(the two upper panels of each $z_i(t)$, from left to right);
$U_{12}=-U$ and $U_{12}=-1.2 \,U$ (the two lower panels of each
$z_i(t)$, from left to right).  We used the initial conditions
$z_1(0)=0.1$, $z_2(0)=0.15$, and $\theta_i(0)=0$. Here, we report
the profiles $|f_{1}|^2$ and $|f_{2}|^2$ as functions of $z$ and
for different values of time, as displayed in the figure. The
quantities $|f_{i}|^2$ are obtained by integrating Eq.
(\ref{1dGPE}) when $U_{12}=-1.2 \, U$. Units are as in Fig. 1 and
Fig. 2.}\label{fig3}
\end{figure}
The condition to find $U^{(II,cr)}_{12}$ is the Eq. (\ref{condition}),
and we obtain that $U^{(II,cr)}_{12}$ coincides with $U^{(I,cr)}_{12}$.

An interesting task, now, is to analyze the influence of the
interaction between the two BECs on the temporal evolution of the
fractional imbalances.  We have fixed $U_{1}$ and $U_{2}$, and
analyzed $z_{1}(t)$ and $z_{2}(t)$ for different values of the
inter-species interaction amplitude $U_{12}$. The greater is the
absolute value of $U_{12}$, the greater is the deformation of the
oscillations around $\langle z(t) \rangle=0$, as shown in the two
upper panels (from left to right)  and in the first lower panel
(from the left) of each $z_{i}(t)$ represented in Fig. 3. Note
that as long as the oscillations are harmonic (see the first panel
of each $z_i(t)$), the time evolution of the fractional imbalances may be described in
terms of a carrier wave of frequency $\omega_{c}$, given by Eq.
(\ref{f1}) with $U_{12}=0$, modulated by a wave of frequency
$\omega_m$. The frequency $\omega_c$ is much greater than
$\omega_m$. We found that there exists a value of the
inter-species interaction amplitude for which the relative
population in each trap oscillates around a non zero time averaged
value, $\langle z(t) \rangle \neq 0$, which corresponds to the
macroscopic quantum self-trapping (MQST) as discussed in
\cite{smerzi} for a single component. To support this
interpretation, we have studied the behavior of the density
profiles of the two species as function of $z$ and for different
values of time. In particular, to find $|f_{i}(z,t)|^2$, we have
numerically solved the two coupled GPE (\ref{1dGPE}) for those
values of the interaction amplitudes for which the self-trapped is
predicted to occur by the coupled pendula equations. We have
summarized the results of this analysis in the last two panels of
Fig. 3.
\begin{figure}
\centerline{\includegraphics[width=8cm,clip]{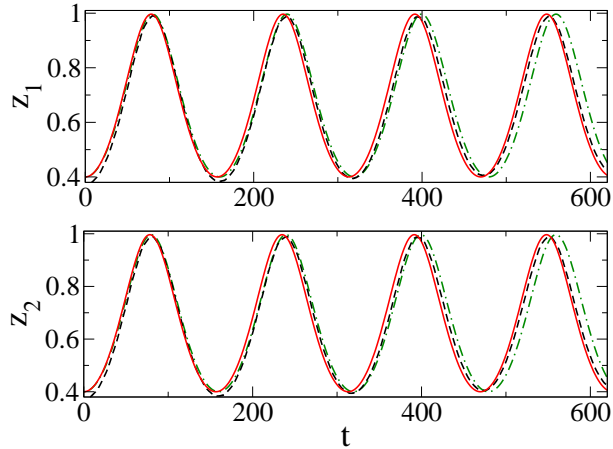}}
\caption{Fractional imbalance of the first and second bosonic vs.
time. The dashed line represents data from integration of Eq.
(\ref{1dGPE}), the dot-dashed line represents solution of Eq.
(\ref{evolution}) with $K_1=K_2 \equiv K=0.0148$, and the
continuous line represents solution of Eq. (\ref{evolution}) with the best-fit
$K$ equal to $0.0151$. Here is $U_{1}=U_{2}=0.1 K$, $U_{12}=-0.14 K$,
$N_1=N_2=100$. We used the initial conditions $z_i(0)=0.4$ and
$\theta_i(0)=0$. The units are as in Fig. 1 and in Fig.
2.}\label{fig4}
\end{figure}
Finally, it is interesting to observe that there is a good
agreement between MQST predicted by the coupled pendula equations
(\ref{evolution}) and the one got by numerically solving the two
1D GPE (\ref{1dGPE}). This comparison is displayed in Fig. 4,
where we can observe oscillation around a non zero time averaged
value of fractional imbalances. We followed the same fitting
procedure adopted in obtaining Fig. 2. \\

At this point, we observe that the question of the AJJ with two
bosonic species could be addressed also from the experimental
point of view. This could be done considering, for example, BECs
binary mixtures of two bosonic isotopes of the same alkali atom
\cite{papp}. The works referenced in \cite{albiez} and \cite{gati}
provide the ideal guide lines for this kind of experiments. By
engineering a double well potential as suggested, for istance, by
Gati {\it et al.} \cite{gati}, the measured fractional imbalances
$z_i(t)$ could be compared with the ones obtained both by solving
the Eq. (\ref{1dGPE}) (see \cite{albiez} and \cite{gati}) and the ODE
system (\ref{evolution}). It could be possible to measure,
moreover, the inter-species $s$-wave scattering length $a_{12}$ by
using, to fix the ideas, the frequency (\ref{f1}). The mixture
could be prepared in such a way that both all the conditions
$z_1(0)=z_2(0) $, $K_1=K_2$, $U_1=U_2$, $N_1=N_2$ - see the
discussion about the four classes (i)-(iv) of stationary points -
are verified and to have small amplitude oscillations around the
stationary point (i). Each $z_i$ oscillates according to the law
$z_i(0)\cos (\omega^{(1)}_{-}t)$ (see Eq. (\ref{coslawi})). Let us suppose to
fix both the characteristic quantities (in our case, they are $\omega_i$, $a$, $z_0$) of
the trapping potential - the group of Heidelberg displayed how this is
possible for a given class of double well traps \cite{albiez}, \cite{gati} - and the intra-species $s$-wave scattering length
$a_{i}$. Then, the functions $\phi_i^{\alpha}$ are known. Then, the Eqs.
(\ref{parameters}) provide the parameters $K_i$ and
$U_i^{\alpha}$.  The measure of the period of $z_i(0)\cos
(\omega^{(1)}_{-}t)$ leads to the corresponding frequency
$\omega^{(1)}_{-}$, i.e. the left-hand side of Eq. (\ref{f1}). The
solution of this equation gives the parameter $U_{12}$. By the
mean of the second line of Eq. (\ref{parameters}) one gets $\tilde
g_{12}$, and, then $a_{12}$.

\section{Conclusions}

We have analyzed the atomic Josephson effect in presence of a
binary mixture of BECs. We have written the Lagrangian of the
system, from which we have derived a system of coupled
differential equations which governs the dynamical behavior of the
fractional imbalance and of the relative phase of each component.
We have analyzed the stable points that preserve the symmetry, and
we have obtained an analytical formula for the frequency
oscillations around these equlibrium points. To this regard, one
of the most interesting features is the possibility to know the
inter-species $s$-wave scattering length from these frequencies.
We have shown that in correspondence of precise values of the
inter-species interaction amplitude, the relative populations
oscillate around a non zero time averaged value. This behavior
corresponds to MQST, a well-known phenomenon when only one
component is taken into account. We have compared the predictions
of GPE with the ones of the coupled pendula equations. We have
performed this comparison in the case of total absence of
interaction, in the case in which only the intra-species
interaction is present, and in the case in which also the
inter-species interaction is involved. We have found that, under
certain conditions, the predictions of GPE agree with those ones
of the coupled pendula equations. 
We have shown that,  under certain hypothesis, it is
possible to obtain analytical expressions 
for the inter-species interaction amplitudes which signs the onset of the
self-trapping. Finally, we have commented about the
possibility to correlate our theoretical work with the experiments
proceeding from the works of the group of Heidelberg, see \cite{albiez} and \cite{gati}  .\\

This work has been partially supported by Fondazione CARIPARO
through the Project 2006: "Guided solitons in matter waves and
optical waves with normal and anomalous dispersion".

\end{document}